\documentstyle[12pt]{article}
\begin{document}
\title{New developments in the theory of heavy quarkonia}
\author{Kacper Zalewski\thanks{Supported in part by the KBN grant
2P302-076-07}\\
Institute of Physics of the Jagellonian University\\ and\\ Institute of
Nuclear Physics, Krak\'ow, Poland}
\maketitle

\abstract{The new approach to decays of heavy quarkonia, based on factorization
and expansions in powers of the relative velocity of the valence quarks is
reviewed.}
\vspace{2cm}

\section{Introduction}
Heavy quarkonia are mesons, where both the valence quarks and the valence
antiquark are heavy i.e. have masses much larger than $\Lambda_{QCD}$. Since
the lifetime of the $t$-quark is too short for hadronization, only the discovery
of the $b\overline{c}(c\overline{b})$ quarkonia is expected besides the
well-known $b\overline{b}$ and $c\overline{c}$ quarkonia. Many features of the
heavy quarkonia can be described by applying to the two-body system $Q +
\overline{Q}$ the ordinary, nonrelativistic Schr\"odinger equation with the
interaction potential depending only on the relative coordinate $r = |
\vec{r}_Q - \vec{r}_{\overline{Q}} |$. The earliest choice for this potential
\cite{EIC}, which is still popular (cf. e.g. \cite{SIS}), is the Cornell
potential

\begin{equation}
V(r) = -\frac{a}{r} + b r + c,
\end{equation}
where $a,b,c$ are positive constants. Choosing the quark mass and solving the
Schr\"odinger equation on finds the energy eigenvalues and the wave functions
corresponding to the bound states.

The energy eigenvalues are interpreted as quarkonium masses. The corresponding
states are labelled like for the hydrogen atom $1S,\;1P,\;2S$ etc. Spin and the
total angular momentum can be included in the usual way by writing
$1\;{}^1S_0$, $1\;{}^3S_1$ etc. We shall not consider states, which are so
heavy that they can decay strongly into pairs of heavy-light mesons containing
the $Q$ and the $\overline{Q}$ separately. Such quarkonia are broad and the
simple description presented here becomes unreliable.

Using textbook formulae and the wave functions one can calculate the
probabilities of the electromagnetic dipole electric and dipole magnetic
transitions between various states of the quarkonia. Higher multipole
transitions could also be calculated, but they have not yet been observed
experimentally. Corrections to the usual formulae are sometimes introduced,
because the ratios of the quarkonium sizes to the wave lengths of the emitted
radiation are much larger than the corresponding ratios in atomic physics.
Similar, but less reliable, calculations can be made for the $n'S \rightarrow
nS$ transitions $(n<n')$ with the emission of a pair of pions.

This approach has two difficulties, which one tries to overcome and a basic
difficulty, which makes people look for completely new approaches. The first
difficulty is that relativistic effects, e.g. fine and hyperfine splittings of
the energy levels, are left out. This is overcome by using various
"relativistic" equations instead of the Schr\"odinger equation and by
introducing more general interaction potentials. Since no generally accepted
two-body relativistic quantum mechanics exists, however, these approaches are
strongly model dependent. They give the necessary splittings, but the centres
of the multiplets are not reproduced any better than in the nonrelativistic
theory. A comparison of 27 relativistic and nonrelativistic potential models
with each other and with experiment is given in the review article \cite{BSK}.
A nonrelativistic potential, which reproduces the data for the
$b\overline{b}$ quarkonia within their experimental errors and also gives good
results for the $c\overline{c}$ quarkonia, has been described in \cite{MZA}.

Another difficulty is that quantum mechanics, strictly speaking, applies to
processes, where the number and quality of particles do not change. A way out
of this difficulty is to assume factorization. E.g. the probability of
annihilation of an $S$-wave quarkonium into a lepton pair is evaluated as a
product of two factors: the probability that $Q$ and $\overline{Q}$ meet, this
is just $|\psi(0)|^2$, and the probability that then they annihilate into
a lepton pair, which is calculated using quantum field theory. Such
calculations are less reliable than the calculations of masses or
non-annihilation transitions, but also they seem to give reasonable agreement
with experiment. Incidentally, the analogy with calculations of decay constants
of the heavy-light mesons (cf. \cite{NZA} and references given there)
suggests that there may be difficulties here, which are hidden for the moment
by other uncertainties.

The main problem with potential models is, however, that their relation to
sound theory is unclear. They also often produce inconsistencies. E.g.
calculating from the Schr\"odinger equation the kinetic energy, one finds
typically $\langle v^2 \rangle \approx 0.08 c^2$ for the
$b\overline{b}(1\;^3S_1)$ and $\langle v^2 \rangle \approx 0.25 c^2$ for the
$c\overline{c}(1\;^3S_1)$. This suggest important relativistic corrections,
which, however, are not contained in the Schr\"odinger equation. Attempts to
calculate radiative
corrections sometimes lead to infinities. One finds anomalous dimensions tending
to infinity when $\frac{v}{c} \rightarrow 0$ i.e. in the nonrelativistic limit
(for a discussion cf. \cite{MSC}) and non-cancelling infrared singularities in
decay amplitudes \cite{BGR}, \cite{BAR1}, \cite{BAR2}.

In the following we present a new approach \cite{MSC}, \cite{BBL}, which
explains at least some of the difficulties of the potential models and also
brings a number of new interesting insights.

\section{Scales}

Let us consider the limit, when the masses of the valence quarks (i.e. of the
valence quark and of the valence antiquark) are large and their velocities
small. Here and in the following velocity means velocity in the rest frame of
the heavy quarkonium. Thus, $v \ll c$, or in the usual system of units, where
the velocity of light $c = 1$, $v \ll 1$. The inequalities

\begin{equation}
M \gg Mv \gg Mv^2 \gg \ldots
\end{equation}
where $M$ is a mass of the order of the mass of the heavy quarkonium, or of the
heavy quark, define a set of well separated energy scales. Let us discuss,
which characteristics of the quarkonia are related to which scale \cite{BBL}.

The largest scale $M$ corresponds to the total mass of the quarkonium. It is
also the inverse of the annihilation radius -- the valence quark and the
valence antiquark can effectively annihilate each other only when they are
separated by a distance of order $\frac{1}{M}$ or smaller. The proposal
\cite{BBL} is to introduce into the theory a cut off $\Lambda = M$ chosen so
that for momenta below $\Lambda$ the motion of the heavy quarks is
non relativistic. The advantage of this choice is that heavy quark production,
hard gluon emission and large relativistic effects in the motion of the heavy
quarks are excluded. The price to pay is that new operators appear in the
Lagrangian. This is like in the theory of electroweak processes, where for low
energy processes a cut off excluding the intermediate bosons can be introduced,
but then four-fermion interactions must be added to the Lagrangian.

The scale $Mv$ is the scale of the momenta of the heavy quarks in the
quarkonium. By the uncertainty principle its inverse is the scale of the
quarkonium radius. The wave vectors of the gluons forming the average
colour field within the quarkonium are also $k \sim Mv$.

The scale $Mv^2$ is the scale of the kinetic energy of the heavy quarks. It is
natural to assume that it is also the scale of the potential energy and of the
excitation
energies of the quarkonia. It happens that $\Lambda_{QCD} \sim 0.4$ GeV is of
the same order of magnitude. It is assumed that the stationary state vectors
of the quarkonia consists not only of components $|Q\overline{Q}\rangle$, where
the valence quarks are in a colour singlet state, but also of components
$|Q\overline{Q},g\rangle$, $|Q\overline{Q},gg\rangle$, \ldots, where the
"dynamic" gluons $g$ have wave vectors $k \sim Mv^2$. Note that in
$|Q\overline{Q},g\rangle$ the $Q\overline{Q}$ system must be in the colour
octet state in order to make the system $Q\overline{Q}g$ a colour singlet.

Lower scales correspond to fine and hyperfine splittings, but we shall not
discuss them here.

Let us see now how these scales can be used to make various useful estimates.
For very heavy valence quarks the radius of the quarkonium is small and it is
believed that the interaction is approximately Coulombic. Since the kinetic
energy and the potential energy should be of the same order, we find

\begin{equation}
\frac{\alpha_s(\frac{1}{R})}{R} \sim Mv^2,
\end{equation}
where $R = \frac{1}{Mv}$ is of the order of the quarkonium radius. Eliminating
$R$ and using the fact that $\alpha_s(\mu)$ is a decreasing function of its
argument we find \cite{BBL}

\begin{equation}
c \sim \alpha_s(Mv) > \alpha_s(M).
\end{equation}
This implies that, at least in the high mass limit, it is inconsistent to
include higher order corrections in $\alpha_s$ (radiative corrections) without
simultaneously including the higher order corrections in $v$ (relativistic
corrections) to at least the same order.

In the Coulomb gauge one finds from standard perturbation theory \cite{BBL}

\begin{equation}
|g\vec{A}| \sim \alpha_s(k) v k.
\end{equation}
This shows that soft gluons couple weakly to slow quarks, even when $\alpha_s$
is not small. It is remarkable that $|g\vec{A}| \sim Mv^3$ both for the gluons
with $k \sim Mv$, for which $\alpha_s(k) \sim v$, and for the much softer
dynamic gluons, for which $k \sim Mv^2$, but $\alpha_s(k) \sim 1$.

Let us estimate now the probability that within the quarkonium the valence
quarks are in a colour octet state, which requires an additional gluon to
compensate the colour charge. It is assumed that this additional gluon is
dynamic. We compare two rough estimates of the quarkonium energy shift due to
the $|Q\overline{Q},g\rangle$ component

\begin{eqnarray}
\Delta_gE \sim P(Q\overline{Q},g) \Delta E,\\
\Delta_gE \sim \langle H|\int d^3x g \vec{v}\cdot\vec{A}|H\rangle.
\end{eqnarray}
The first estimate, where $\Delta E \sim Mv^2$ is an estimate of the energy of
the additional gluon, is probabilistic. In the second, perturbative estimate,
where $H$ denotes the quarkonium, $\vec{v}\cdot\vec{A} \sim Mv^4$ and the other
factors do not change this estimate of $\Delta_gE$. Eliminating $\Delta_gE$ we
find \cite{BBL}

\begin{equation}
P(Q\overline{Q}g) \sim v^2.
\end{equation}
Thus, contrary to the picture suggested by potential models, this probability is
small, but nonzero.

Let us note one more estimate. Since the squared modulus of the wave function
integrated over the volume occupied by the quarkonium ($\sim (Mv)^{-3}$)
gives unity, $\psi \sim M^{\frac{1}{2}} v^{\frac{3}{2}}$ in the region of
interest.

\section{Lagrangian}

The full Lagrangian for the quarkonium is, of course, the ordinary Lagrangian
of the standard model. Here, however, we need an effective Lagrangian with a
cut off at $\Lambda = M$. The price for the cut off is that the Lagrangian has
an infinite number of terms. In the small velocity limit, however, only a few
of these terms survive. Systematic expansions in powers of velocity for the
necessary matrix elements can be constructed by adding at each step only a
finite number of terms to the previous approximation to the effective
Lagrangian. In general the effective Lagrangian is written in the form

\begin{equation}
{\cal L} = {\cal L}_{light} + {\cal L}_{heavy} + \delta{\cal L}.
\end{equation}
Here

\begin{equation}
{\cal L}_{light} = -\frac{1}{2}TrG_{\mu\nu}^aG^{\mu\nu a} + \sum_j
\overline{q}_j iD\!\!\!\!\!\;/ q_j
\end{equation}
is the Lagrangian for the gluons and for the light quarks assumed for
simplicity massless. The second term

\begin{equation}
{\cal L}_{heavy} = \psi^{\dag}\left( iD_t + \frac{\vec{D}^2}{2M}\right)\psi +
\chi^{\dag}\left( iD_t - \frac{\vec{D}^2}{2M}\right)\chi
\end{equation}
is the nonrelativistic approximation to the Lagrangian of the heavy quarks
(created by $\psi^{\dag}$) and heavy antiquarks (created by $\chi$) interacting
with the colour field. The mass $M$, which till now has been only esimated, is
unambiguously defined by this equation. Everything else necessary to make the
effective theory equivalent to the exact theory at the low energy scale is
contained in the correction term $\delta{\cal L}$. The idea is to classify the
infinitely many terms contributing to $\delta{\cal L}$ according to their order
with respect to velocity. Using information about the scales of various
operators, as described in the previous section, one sees that ${\cal
L}_{heavy}$ contains terms of order $v^5$. Putting $\delta{\cal L} = 0$, one
obtains an approximation, where in particular the numbers of heavy quarks and
heavy antiquarks are separately conserved. This approximation may be used to
estimate the masses of the quarkonia, but not their annihilations into light
particles. The next approximation includes the bilinear operators of order
$v^7$

\begin{equation}
\delta_2{\cal L} = \frac{c_1}{8M^3}\left[\psi^{\dag}(\vec{D}^2)^2\psi\right.
- \left.\chi^{\dag}(\vec{D}^2)^2\chi\right] + \cdots,
\end{equation}
where the dots replace the other terms with the same structure in the heavy
fields. Three of them are of the same order $v^7$. The coefficients $c_i$ are
calculable and made dimensionless by explicitly factoring out suitable powers
of $M$. This approximation gives an improved description of the structure of
the quarkonia, but again does not allow the annihilation of heavy quarks.

In order to describe annihilation one needs four-heavy-fermion operators, which
occur in the next order in the expansion of $\delta{\cal L}$. One has

\begin{equation}
\delta_4{\cal L} = M^{-2} \sum_{j=1}^4 f_j \psi^{\dag} O_j \chi \chi^{\dag} O_j
\psi + \cdots,
\end{equation}
where $O_1 = 1,\;O_2 = \vec{\sigma},\;O_3 = \lambda^a,\;O_4 = \lambda_a
\vec{\sigma}$. The $f_j$ are calculable, dimensionless coefficients and the
dots denote higher dimension operators with the same structure in heavy quark
fields. The operators occurring in this formula are of order $v^6$, but the
coefficients $f_j$ are of order $\alpha_s^2$, therefore, $\delta_4{\cal L}$ is
of order $v^8$. These terms in the Lagrangian make annihilation possible.
Actually, the first term annihilates the $^1S_0$ colour singlet states, the
second the $^3S_1$ colour singlet states, the third the colour octet $^1S_0$
states and the fourth the colour octet $^3S_1$ states. For annihilation of
quarkonium $H$ into light hadrons one finds

\begin{equation}
\Gamma(H \rightarrow l.h.) = 2 Im\langle H|\delta_4{\cal L}|H\rangle,
\end{equation}
or approximately

\begin{equation}
\Gamma(H \rightarrow l.h.) = 2 M^{-2} \sum_{j=1}^4 Im f_i\; \langle
H|\psi^{\dag} O_j \chi \chi^{\dag} O_j \psi |H\rangle.
\end{equation}
Note that the operators $j=3,4$ can annihilate only the quarkonium component
$|Q\overline{Q},g\rangle$, which has an amplitude of the order
$\sqrt{P(Q\overline{Q},g)} \sim v$, and that that reduces the contribution to
$\Gamma$ by a factor of order $v^2$. This usually justifies the omission of
$j=3,4$ in low order calculations, but in the following section we show an
example, where this reduction is compensated by another factor and the
inclusion of $j=3,4$ is crucial.

For annihilations into leptons and/or photons the final state is a vacuum of
hadrons, therefore, the operators $j=3,4$ cannot contribute and, moreover, e.g.

\begin{equation}
\langle H| \psi^{\dag} \chi \chi^{\dag} \psi |H\rangle = |\langle
H|\psi^{\dag}\chi|0\rangle|^2.
\end{equation}
The same formula used for decays into light hadrons is only an approximation
known as the vacuum saturation. Rigorously

\begin{equation}
\langle  H|\psi^{\dag} \chi \chi^{\dag} \psi |H \rangle = \sum_X |\langle
H|\psi^{\dag} \chi |X\rangle|^2,
\end{equation}
where the vectors $|X\rangle$ are a complete set of states. Then the vacuum
saturation means neglecting all the contributions with $|X\rangle \neq
|0\rangle $. Let us estimate the terms, which are thus neglected. Terms with
$|X\rangle = |g\rangle$ do not contribute, because the operator
$\psi^{\dag}\chi$ does not change the colour and the state $|H\rangle$ is by
assumption colourless. Terms with $|X\rangle = |gg\rangle$ contain only the
components of $|H\rangle$, which besides $Q\overline{Q}$ contain two dynamic
gluons. Such components of $H$ are of order $v^2$, therefore, the corresponding
terms on the right hand side of the previous formula are reduced by a factor of
order $O(v^4)$. For $|X\rangle = |q\overline{q}\rangle$ or $|X\rangle =
|ggg\rangle$ the reduction is even stronger. Thus, in the present approach the
error inherent in the vacuum saturation approximation can be rigorously
estimated and is small.

\section{Examples}

As our first example let us consider the annihilation of $\eta_c$ into light
hadrons. The leading term is

\begin{equation}
\Gamma(\eta_c \rightarrow l.h.) = \frac{Im\,f_1}{M^2}
\langle\eta_c|\psi^{\dag} \chi \chi^{\dag} \psi|\eta_c\rangle.
\end{equation}
Using the vacuum saturation approximation as explained above

\begin{equation}
\langle\eta_c|\psi^{\dag} \chi \chi^{\dag} \psi|\eta_c\rangle = |\langle
\eta_c|\psi^{\dag} \chi |0\rangle|^2.
\end{equation}
The matrix element on the right-hand side can be related to a radial wave
function averaged over a volume of order $M^{-3}$

\begin{equation}
|\langle \eta_c|\psi^{\dag} \chi |0\rangle| = \sqrt{\frac{3}{2\pi}}
|\overline{R}_{\eta_c}(0)|^2,
\end{equation}
where the factor three under the square root is the number of colours. Using
again the scale estimates one finds

\begin{equation}
\overline{R}_\psi(0) = \overline{R}_{\eta_c}(0)(1 + O(v^2)).
\end{equation}
Thus, up to corrections of order $v^2$ the decays of corresponding $^1S_0$ and
$^3S_1$ states are governed by the same matrix element. This is the realization
in the present approach of the spin symmetry from the heavy quark effective
theory. At the same level of precision one can connect the matrix elements
for the production and for the decays of the heavy quarkonia \cite{BBL}.

Let us note an advantage of the present approach as compared with potential
models. Certain expressions, which are infinite in potential models, like
$\vec{\nabla}^2\,R(r)$ for $r \rightarrow 0$, can be here regularized by
standard methods of quantum field theory.

As our second example let us consider the decay of a $P$-wave quarkonium into
light hadrons. In potential models this problem was plagued by infinities. We
shall show how it is resolved here. Let us note first that for $P$-states
$\psi(0) = 0$. Therefore, choosing for definiteness the $1\:^1P_1$ quarkonium
$h_c$ as our example and identifying the wave function $\overline{R}_{h_c}$
with the wave function given by potential models we have

\begin{equation}
\langle h_c|\psi^{\dag}\chi\chi^{\dag}\psi|h_c\rangle = 0.
\end{equation}
The next term contains

\begin{equation}
\label{matrDD}
\langle h_c|\psi^{\dag}(-\frac{i}{2}\stackrel{\leftrightarrow}{D}) \chi
\chi^{\dag}((-\frac{i}{2}\stackrel{\leftrightarrow}{D})\psi|h_c\rangle =
\frac{9}{2\pi} |\overline{R}_{h_c}'(0)|^2,
\end{equation}
where $R_{h_c}'$ is the derivative with respect to $r$ of the radial wave
function of the valence quarks in the quarkonium $h_c$. The right-hand side is
well known from potential models. This matrix element, however, is suppressed
by two powers of $v$ as compared to the matrix element responsible for the
decays of $S$-wave quarkonia into light hadrons. At the same order in $v$ there
is another matrix element

\begin{equation}
\label{matcol}
\langle h_c| \psi^{\dag} \lambda^a \chi \chi^{\dag} \lambda^a \psi |h_c\rangle,
\end{equation}
which corresponds to the annihilation of the $|Q\overline{Q},g\rangle$
component of $h_c$ and which has no analogue in potential models. The
suppression by $v^2$ due to the choice of the $|Q\overline{Q},g\rangle$
component of the $h_c$ state vector is compensated by removing the factor of
order $v^2$ due in the matrix element (\ref{matrDD}) to the two factors
$\stackrel{\leftrightarrow}{D}$. To put it more intuitively, the gain due to
the fact that the colour octet $Q\overline{Q}$ system is in an $S$-state and
the quark and antiquark can easily meet to annihilate compensates the penalty
for introducing the additional dynamic gluon. When both terms (\ref{matrDD})
and (\ref{matcol}) are included, one finds that the infrared divergencies of
the radiative corrections cancel and thus a meaningful calculation can be
performed. One can also use the renormalization group equations to improve the
results \cite{BBL}.

\section{Conclusions}

The problem of heavy quarkonia is difficult. Brute force expansions in
$\alpha_s$ or in $\frac{1}{M}$ are unlikely to work. Simple-minded potential
models work well within their applicability range, but the reason for their
success is not understood and obvious contradictions are encountered on the
way. Also the parameters of these models (e.g. quark masses) have no
definitions sufficiently precise to connect them to quantities occurring in QCD
(in the case of quark masses to $\overline{MS}$ or pole masses).

A recently proposed combination of factorization and expansions of parts of the
matrix elements in powers of the relative velocity $v$ \cite{MSC} \cite{BBL}
seems very promising. It is much closer to rigorous QCD than the potential
models. Its various assumptions and approximations can be rigorously studied,
because the relevant quantities are defined in a way, which is understandable
from the point of view of QCD. Its weakness at present is that it introduces
many matrix elements, which in principle are calculable from lattice QCD, but
in practice may remain for many years to come just phenomenological parameters.
It should be stressed, however, that the theory already has given some
interesting approximate relations between these matrix elements and that it
removes the inconsistencies, which have been plaguing the potential models.

\end{document}